# Spin dependent electron-phonon interaction in SmFeAsO by low-temperature Raman spectroscopy

L. Zhang[1,¶], P. F. Guan[1,¶], D. L. Feng[2], X. H. Chen[3], S. S. Xie[4], M. W. Chen[1]*

[1] WPI Advanced Institute for Materials Research, Tohoku University, Sendai 980-8577, Japan

[2] Department of Physics, Fudan University, Shanghai, China

[3] Department of Physics, University of Science and Technology of China, Hefei 230026, China

[4] Institute of Physics, China Academy of Sciences, Beijing, China

## Abstract

**The interplay between spin dynamics and lattice vibration has been suggested as an important part of the puzzle of high-temperature superconductivity. Here we report the strong interaction between spin fluctuation and phonon in SmFeAsO, a parent compound of the iron arsenide family of superconductors, revealed by low-temperature Raman spectroscopy. Anomalous zone-boundary-phonon Raman scattering from spin superstructure was observed at temperatures below the antiferromagnetic ordering point, which offers compelling evidence on spin dependent electron-phonon coupling in pnictides.**

\* E-mail: mwchen@wpi-aimr.tohoku.ac.jp

¶ These authors contributed equally to this work

**Introduction**

The newly discovered FeAs-based superconductors[1-3] have renewed much experimental and theoretical enthusiasms on high-temperature superconductivity. Much effort has been devoted to investigate the electron pair interaction which permits the formation of the superconductivity.[4,5] For conventional superconductors, this interaction can be provided by the exchange of phonon, but the theory[6-8] indicates the simple "electron-phonon interaction" in the pnictides is too weak to account for the high critical transformation temperatures ($T_c$). Recently, inelastic neutron scattering and isotope effect experiments as well as theoretical calculations suggest that spin fluctuation may be a possible approach to provide the energy for electron-pair formation and the interplay between the spin dynamics and lattice vibrations may be the key to understand the mechanism of high $T_c$ superconductivity.[9-14] Despite low-temperature phonon splitting and softening that may be associated with spin-phonon interaction have been observed by neutron scattering, inelastic x-ray scattering and Raman studies,[11,14-16] the unambiguous evidence on the spin dependent electron-phonon interaction has not been achieved in pnictides.

In this report, the evolution of phonon spectra of $SmFeAsO_{1-x}F_x$ (x=0, 0.2) with temperature was investigated by low-temperature Raman spectroscopy combining with density functional theory (DFT) calculations. An anomalous zone-boundary-phonon mode from spin superstructure was found in SmFeAsO at temperature below the antiferromagnetic ordering point ($T_{SDW}$), which offers compelling evidence on spin dependent electron-phonon coupling in pnictides.

**Experimental Section**

**Sample preparation.** Polycrystalline samples with nominal compositions of $SmFeAsO_{1-x}F_x$ (x=0, 0.2) were synthesized by conventional solid state reaction using high purity SmAs, $SmF_3$, Fe, and $Fe_2O_3$ as green powders.[17] $SmFeAsO_{0.8}F_{0.2}$ has been demonstrated to be a superconductor with $T_c$ of ~55K ( see **Fig. S1** in the Supporting Information).[18]

**Raman measurements.** A micro-Raman spectrometer (Renishaw InVia RM 1000) with incident wavelengths of 514.5 nm and 632.8nm was used in this study. The sample surfaces were carefully polished for Raman characterization. The probe laser beam of ~3μm in diameter is smaller than the grain sizes of the polycrystalline samples (*see* **Fig. S2** in the Supporting Information), which allows us to acquire single phase spectra by selecting detection sites. To avoid possible sample damage by laser irradiation, the laser power was kept to a minimum value.[19] Each spectrum pattern presented in this Letter was averaged over a number of spectra collected at different points with same acquisition conditions in an un-polarized mode. Low-temperature Raman studies were conducted using a heating/cooling stage where liquid nitrogen (LN) was used as a cryogenic source. The sample temperature was measured using a Pt-based thermal couple with an accuracy of ±0.1 K.

**Phonon calculation.** Phonon frequencies of SmFeAsO were calculated by a linear-response method based on the framework of the density functional theory, performed within the pseudopotential approximation for the electron-ion interaction.[18,] A local-density approximation with gradient corrections was used for the exchange correlation energy [21] and the phonon dispersion was calculated based on the density-functional perturbation theory.[22] The self-consistent calculation was accomplished using experimental lattice parameters and energy minimized internal atomic positions.

**Results and Discussion**

As suggested by neutron and synchrotron powder diffraction, the SmFeAsO compound undergoes a structural phase transition from tetragonal (space group P4/nmm) to orthorhombic (space group *Cmma*) structures just above the temperature ($T_{SDW}$) of antiferromagnetic (AFM) ordering during cooling, and fluorine doping suppresses the structural and magnetic transitions.[23] We investigate the temperature dependence of the phonon modes of SmFeAsO and SmFeAsO$_{0.8}$F$_{0.2}$ using low-temperature Raman spectroscopy. As shown in **Fig. 1(a)**, the SmFeAsO and SmFeAsO$_{0.8}$F$_{0.2}$ have similar phonon spectra at temperatures above the structure transition and most of the phonon modes are in agreement with the theoretical calculation of the nonmagnetic *P4/nmm* structure (see **Table 1**) and previous experimental observations.[24, 25] The main Raman-active phonon modes of the compounds can be categorized into two groups. One includes four zone-centered phonons that arise from axial motion of Sm ($A_{1g}$, 171cm$^{-1}$), As ($A_{1g}$, 201cm$^{-1}$), Fe ($B_{1g}$, 210 cm$^{-1}$) and O ($B_{1g}$, 347 cm$^{-1}$). The other is from the in plane modes with $E_g$ symmetry at 119 cm$^{-1}$ (Sm), 132 cm$^{-1}$ (As), and 270 cm$^{-1}$ (As & Fe). When temperature decreases, the structure of the compounds become more compact with the reduction of bond distance, leading to a shift of Raman bands to higher frequencies. In this study, the Fe modes show a slightly more pronounced hardening than the As ones upon cooling, which results in a splitting of the As and Fe modes at ~270 cm$^{-1}$. Moreover, the temperature dependence of the Raman scattering of the SmFeAsO$_{1-x}$F$_x$ (x=0.0, 0.2) compounds is anisotropic and the characteristic modes display different temperature dependences in both phonon energy and phonon linewidth (full width at half-maxima, FWHM). As shown in **Fig. 1 (b)** and **(c)**, with temperature decreasing the Raman modes of the SmFeAsO$_{0.8}$F$_{0.2}$ compound are stiffened while the FWHM near monotonously decreases. In contrast, anomalous variation in both phonon energy and FWHM of SmFeAsO can be observed at the temperature of ~150 K that corresponds to the structural phase transition. It has been reported that in the SmFeAsO$_{1-x}$F$_x$ system with x<0.15 the tetragonal structure is initially robust upon cooling showing a normal contraction of the lattice parameters. At the structural phase transition temperature, the lattice constant *a* splits due to the transformation from tetragonal to orthorhombic structure.[23] With the structural phase transition all the in-plane phonons split into two non-degenerated phonons as listed in **Table 1**. But the splitting is apparently too small [26] to be identified in our experiment and only leads to an anomalous increasing in the peak width (**Fig.1 (c)**). In contrast, since the structural phase transition is suppressed by fluorine doping, the FWHM of phonon modes of SmFeAsO$_{0.8}$F$_{0.2}$ decreases normally with temperature (**Fig.1 (c)**).

It is worth noting that a new vibrational mode of SmFeAsO appears at ~224cm$^{-1}$ (marked with arrows) when temperatures are below 130K (**Fig.1 (a)**). The intensity of this anomalous Raman band noticeably increases with temperature decreasing and at the temperature of ~78 K turns to be as high as those of the normal phonon modes of the compound, whereas this mode is absent in the SmFeAsO$_{0.8}$F$_{0.2}$ compound. The new peak (R$^1$) can be observed under both 514.5nm and 632.8nm laser excitations as shown in **Fig. 2(a)**, verifying that this mode is not due to resonant Raman scattering. Since the structure distortion caused by the structural phase transition from the *P4/nmm* to *Cmma* phases does not introduce any new Raman-active mode (see **Table I**) but just leads to the band broadening due to the splitting of E$_g$ modes, the anomalous Raman band R$^1$ is most likely associated with the AFM transition that takes place at a temperature of ~130 K.[9] We measured the amplitude evolution of the new Raman band with temperature (**Fig. 2(b)**). Consistent with the temperature dependence of AFM ordering in LaOFeAs measured by neutron scattering, the evolution of the Raman intensity of the R$^1$ band can be fitted by,[27]

$$\sigma \propto \left| R + M \langle S(\vec{r}_i) S(\vec{r}_j) \rangle / S^2 \right|^2 \quad (1)$$

where both R and $M \langle S(\vec{r}_i) S(\vec{r}_j) \rangle / S^2$ are the spin-dependent term, and $S(\vec{r}_i) S(\vec{r}_j)$ can simply be modeled with a mean field description for magnetization with T$_{SDW}$ of ~131 K.[9] The direct link between the Raman peak amplitude and the order parameter of the AFM transition further suggests that the anomalous Raman band R$^1$ originates from the AFM ordering. There are two possible explanations on the AFM related phonon mode: one is light scattering by spin-density wave and the other is spin-dependent phonon scattering. Since magnon scattering usually has a relatively smaller intensity and a larger FWHM compared with normal phonon modes,[5,28-30] the strong R$^1$ Raman band with the phonon energy of ~ 28meV, narrow FWHM of ~0.8meV and high intensity cannot be explained by the excitation of magnons alone. Therefore, this new peak appears to arise from the spin-dependent phonon Raman scattering.

To understand the possible spin-dependent phonon Raman scattering, we calculated the phonon dispersion curves of SmFeAsO with a *Cmma* structure (**Fig. 3(a)**). All the phonon modes at the point Γ of the dispersion curves can be found in the experimental spectrum and $E_g$ modes have a tendency to split into two due to the slight lattice distortion of the structural phase transition. However, at the Γ point, the phonon mode corresponding to the new Raman peak $R^1$ (224 cm$^{-1}$) cannot be found. Instead, the frequency of a phonon mode at M point (marked with $M^1$) matches well with that of the new Raman band $R^1$ in **Fig. 2**. Since a Raman active phonon mode must satisfy the condition of vanishing wave vector $\vec{q}_R = 0$, in order to be Raman active the $M^1$ phonon need to be folded into the zone centre. This can be produced by the interaction between the spin system and phonons, via the phonon-induced modulation of the exchange interaction, provided the magnetic cell is larger than the crystallographic cell. In this process, spin-dependent electron-phonon interaction will play a key role in the phonon folding.[31-33] Concerning the phonon modulation of the exchange energy, the spin-dependent electron-phonon (EP) interaction can be written as follows:[31]

$$H_{EP} = (2N)^{-1} \sum_{\vec{q}\lambda} Q_{\vec{q}\lambda}(t) \sum_{m} e^{i\vec{q}\cdot\vec{r}_m} \sum_{i\neq j} [\sum_{n} \vec{w}(k|\vec{q}\lambda) \cdot (\partial J_{ij}/\partial \vec{r}_{mn})] \vec{S}_i \cdot \vec{S}_j \quad (2)$$

where $Q_{\vec{q}\lambda}(t)$ and $\vec{w}(k|\vec{q}\lambda)$ are the normal-coordinate operator and the polarization vector, respectively, for wavevector $\vec{q}$ and branch index $\lambda$; $\vec{r}_{mn} = \vec{r}_m + \vec{r}_n$ is the position of the $m^{th}$ atom in the $n^{th}$ crystallographic cell ($m=1,2,...,N$). $J_{ij}$ is the exchange constant between spins $i$ and $j$ in the magnetic cell; $\partial J_{ij}/\partial \vec{r}_{mn} \equiv \hat{i}\partial J_{ij}/\partial x_{mn} + \hat{j}\partial J_{ij}/\partial y_{mn} + \vec{k}\partial J_{ij}/\partial z_{mn}$ and $\vec{S}_i$, $\vec{S}_j$ are the spin operators of the two electrons $i$ and $j$. Considering the periodicity of the magnetic lattice, the $H_{EP}$ can be rewritten as :

$$H_{EP} = \sum_{\vec{g}_m}\sum_{\vec{q}\lambda} Q_{\vec{q}\lambda}(t)\varepsilon_{-\vec{g}_m}(\vec{q}\lambda)\delta(\vec{q}-\vec{g}_m) \tag{3}$$

where $\delta(\vec{q}-\vec{g}_m)$ is the Dirac delta function, $\vec{g}_m$ is the reciprocal vectors of the magnetic lattice and $\varepsilon_{-\vec{g}_m}(\vec{q}\lambda)$ contains the information on the spin-phonon coupling that can be determined by inspecting the spin arrangement (magnetic moment fluctuation). Based on equation (3), a complete link between the folding phonons and the spin structure can be made since the selection principle of the folding phonons is based on two conditions: 1) $\vec{q}_R = \vec{q} - \vec{g}_m = 0$, i.e. $\vec{q} = \vec{g}_m$, due to the magnetic ordering; and 2) $\varepsilon_{-\vec{g}_m}(\vec{q}\lambda) \neq 0$ with $\vec{q} = \vec{g}_m$ due to spin-phonon interaction. If $H_{EP} \neq 0$, both conditions must be satisfied simultaneously.

Base on the magnetic and crystal structures of SmFeAsO at low temperatures, a primitive magnetic cell can be defined by the vectors $\vec{a}_m = \vec{a}+\vec{b}$, $\vec{b}_m = \vec{b}-\vec{a}$ and $\vec{c}_m = \lambda\vec{c}$, where $\vec{a},\vec{b},\vec{c}$ refer to the vectors of the crystallographic primitive cell and $|\vec{a}|=|\vec{b}|$. **Fig. 3(b)** represents a projection of crystallographic and magnetic primitive cells onto (001) plane with Fe atoms and As atoms denoted by red and purple balls. The crystallographic primitive cell is marked with blue line whereas the magnetic primitive cell is plotted with yellow line. The Brillouin zones corresponding to crystallographic and magnetic primitive cells are shown in **Fig. 3(c)**. It is obvious that the wave vector $\vec{q}$ at the M point ($\vec{q}_M = (\pi/a, \pi/b, 0)$) of the crystallographic zone lies at the center of the magnetic zone $\Gamma_m(\vec{g}_m)$. Thus, the phonon modes at the M point ($\vec{q} = \vec{g}_m$) are expected to fold into the centre of the crystallographic Brillouin zone. When we check the schematic displacement patterns of the phonon modes that belong to the M point, it can be found that only the mode $M^1$ can bring an obvious perturbation to the spin arrangement because it originates from the vibration of Fe atoms. The schematic displacement patterns of the phonon mode $M^1$ is shown in **Fig. 3(d)**. The vibration of Fe atoms will change their short-range exchange integrals, which can lead to magnetic moment fluctuation and thereby the two conditions, $\varepsilon_{-\vec{g}_m}(\vec{q}\lambda) \neq 0$ at $\vec{q} = \vec{g}_m$, for spin-phonon interaction can be satisfied at the $M^1$ point. Consequently, the appearance of the strong Raman mode $R^1$, corresponding to the folded $M^1$ phonon mode, provides compelling evidence on the spin dependent electron-phonon coupling in SmFeAsO at temperatures below $T_{SDW}$.

**Conclusion**

Based on the AFM ground state structure of ReFeAsO at temperatures below $T_{SDW}$, the low-temperature phonon mode of SmFeAsO at ~224 cm$^{-1}$ can be unambiguously explained by the spin dependent electron-phonon coupling, rather than by a structural phase transition. The absence of the $R^1$ modes in SmFeAsO$_{0.8}$F$_{0.2}$, demonstrates that the spin-dependent phonon Raman scattering is suppressed, accompanying with the frustration of magnetic ordering by fluorine doping. This further supports the AFM origin of the anomalous $R^1$ Raman band of SmFeAsO.

**Acknowledgements** We thank Xiaofeng Jin, S. Maekawa and Takeshi Egami for constructive discussion. This work was sponsored by "Global COE for Materials Research and Education" and "World Premier International Research Center (WPI) Initiative" by the MEXT, Japan.

*Table1*. The main optical phonons of SmFeAsO with tetragonal *(P4/nmm)* and orthorhombic *(Cmma)* structures. The parameters used in the calculations are listed in the Table S1 of Supporting Information. [100], [010] and [001] in the vibration column are parallel to *a*, *b*, *c* crystal axes, respectively.

| Experimental (cm$^{-1}$) | | Calculated (cm$^{-1}$) | | Symmetry | Active | Atom | Vibration |
|---|---|---|---|---|---|---|---|
| 300K | 90K | *P4/nmm* | *Cmma* | | | | |
| | | 67 | 64 (70) | $E_u$ | IR | Re,Fe,As | [100] or [010] |
| | | 96 | 93 | $A_{2u}$ | IR | Re,Fe,As | [001] |
| 119 | 120 | 118 | 116 (120) | $E_g$ | Raman | Re | [100] or [010] out-of-phase |
| 134 | 139 | 146 | 144 (146) | $E_g$ | Raman | As, Fe | [100] or [010] out-of-phase |
| 170 | 175 | 175 | 174 | $A_{1g}$ | Raman | Re | [001] out-of-phase |
| 202 | 203 | 208 | 199 | $A_{1g}$ | Raman | As | [001] out-of-phase |
| 210 | 215 | 215 | 214 | $B_{1g}$ | Raman | Fe | [001] out-of-phase |
| | 225 | | | | | | |
| | | 255 | 247 | $A_{2u}$ | IR | Fe, As | [001] |
| 269 | 276 | 282 | 281 (283) | $E_g$ | Raman | Fe | [100] or [010] out-of-phase |
| | 303 | 301 | 297 (304) | $E_{2u}$ | IR | Fe, As | [100] or [010] |
| 347 | 351 | 348 | 349 | $B_{1g}$ | Raman | O | [001] out-of-phase |
| | | 363 | 360 (365) | $E_u$ | IR | Re, O | [100] or [010] in-phase |
| | | 431 | 431 | $A_{2u}$ | IR | Re, O | [001] in-phase |
| 503 | 510 | 506 | 496 (515) | $E_g$ | Raman | O | [100] or [010] out-of-phase |

**Figure captions**

**Figure 1**. Low-temperature Raman spectroscopy of SmFeAsO and SmFeAsO$_{0.8}$F$_{0.2}$. **(a)** Micro-Raman spectra of SmFeAsO and SmFeAsO$_{0.8}$F$_{0.2}$ at various temperatures. (The new peak appears upon cooling is marked with arrowheads) **(b)** Temperature dependence of phonon energy; and **(c)** temperature-dependence of phonon FWHM of in-plane mode for Fe&As.

**Figure 2**. Evolution of the anomalous phonon mode with temperature. **(a)** Micro-Raman spectra of SmFeAsO at room temperature and at 90K with a 514.5 nm excitation. The insert shows the spectrum at 90K with a 632.8nm excitation. The new peak at ~224 cm$^{-1}$ is noted by R$^1$ with a red line. **(b)** Temperature dependence of the intensity of the new phonon. The relative intensity of the new phonon is normalized by the intensity of the strongest peak at ~170 cm$^{-1}$. The solid line is a simple fit based on mean field theory.

**Figure 3.** Theoretical analysis of spin-dependent phonon Raman scattering of SmFeAsO. (a) Calculated phonon dispersion curves of SmFeAsO with a *Cmma* structure for $\Gamma(0,0,0)$, M $(\pi/2a, \pi/2b, 0)$, Z $(0, 0, \pi/2c)$ and A $(\pi/2a, \pi/2b, \pi/2c)$ The M$^1$ mode is in good agreement with the new Raman band (R$^1$) in **Fig. 2**. (b) Projected crystallographic (magnetic) primitive cells are shown by blue (yellow) lines, in which only the Fe (red) and As (purple) atoms are displayed. The Fe spins are drawn in white and display a stripe-like anti-ferromagnetic order. **(c)** Simply Brillouin zone of FeAs-based superconductor system (such as SmFeAsO) corresponding to the nonmagnetic unit cell with a *Cmma* structure (red line) and the stripe-like

anti-ferromagnetic phase (blue line). **(d)** Projected displacement patterns for the $M^1$ phonon mode (yellow arrow). The white line shows the unit cell of the *Cmma* structure and the exchange integrals are also depicted.

**Figure 1**.

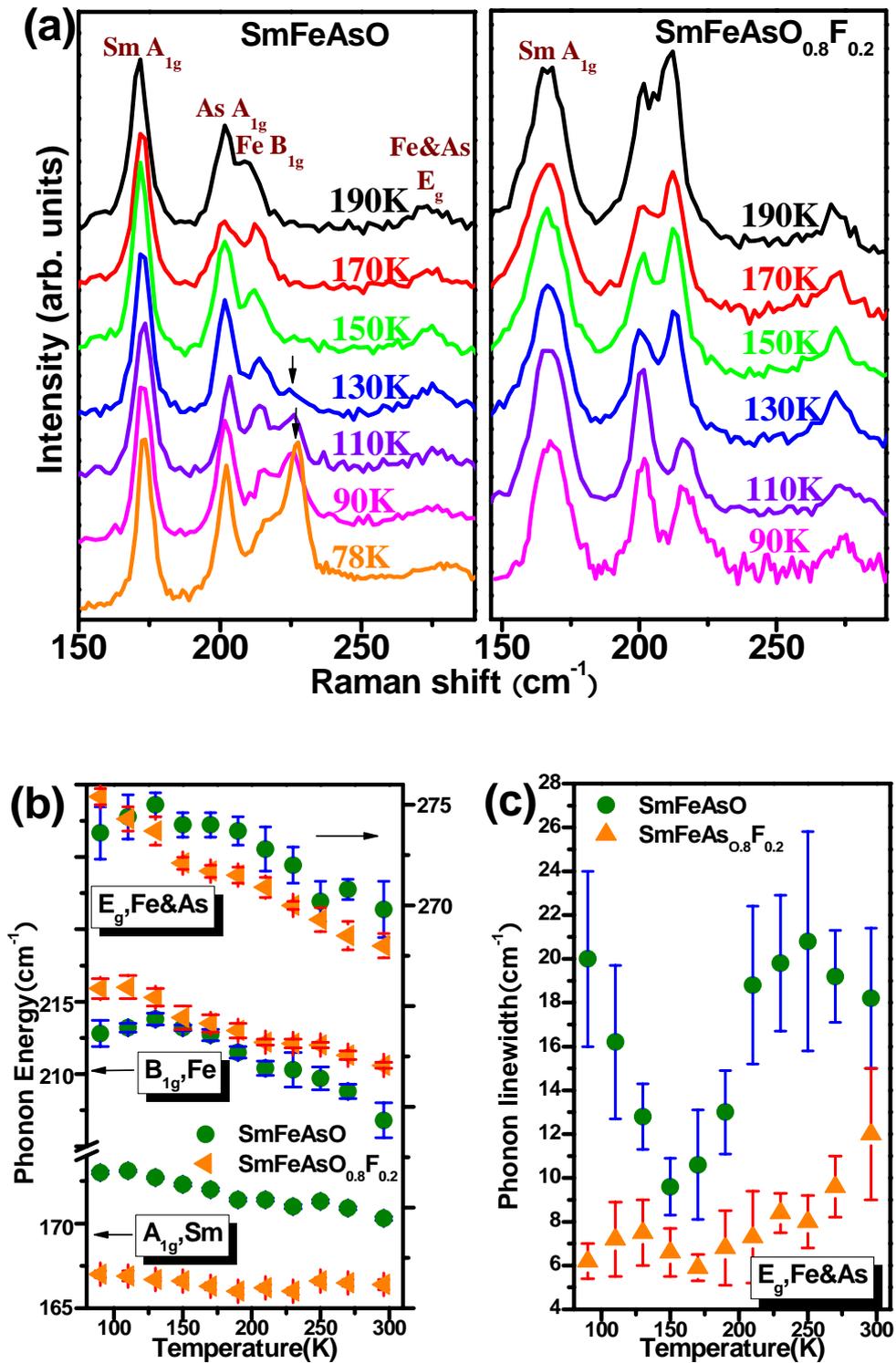

**Figure 2**.

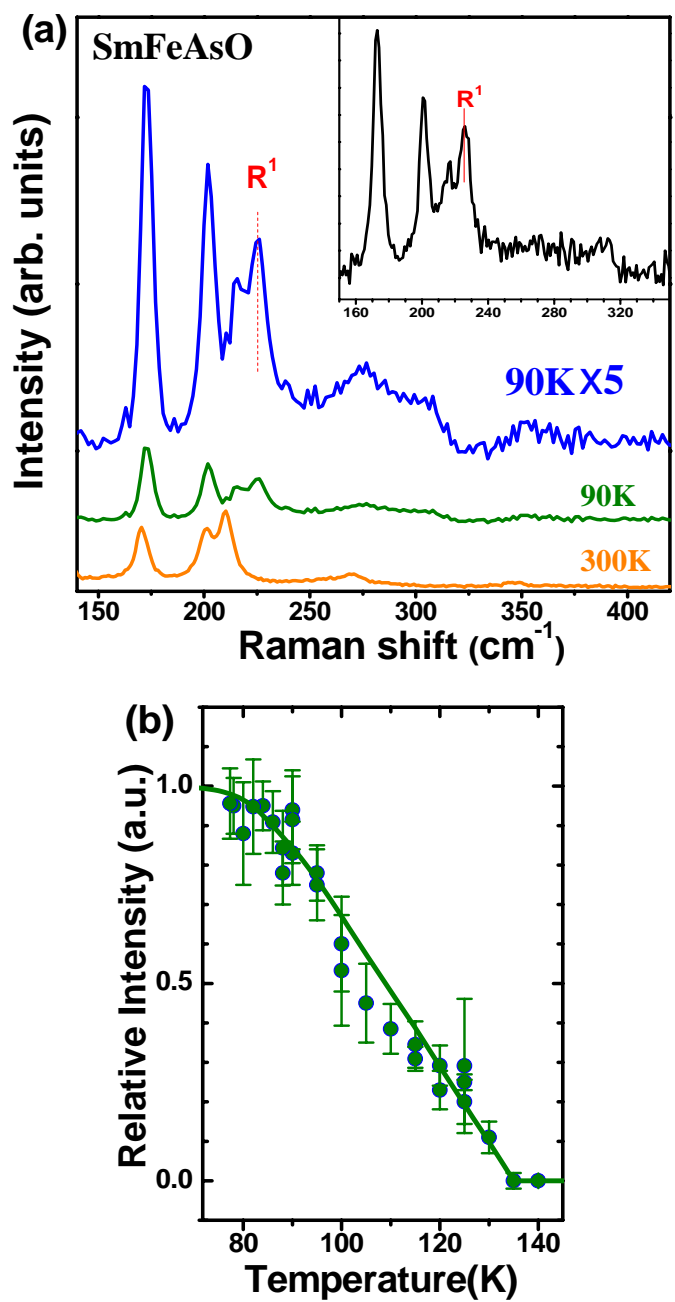

Figure 3.

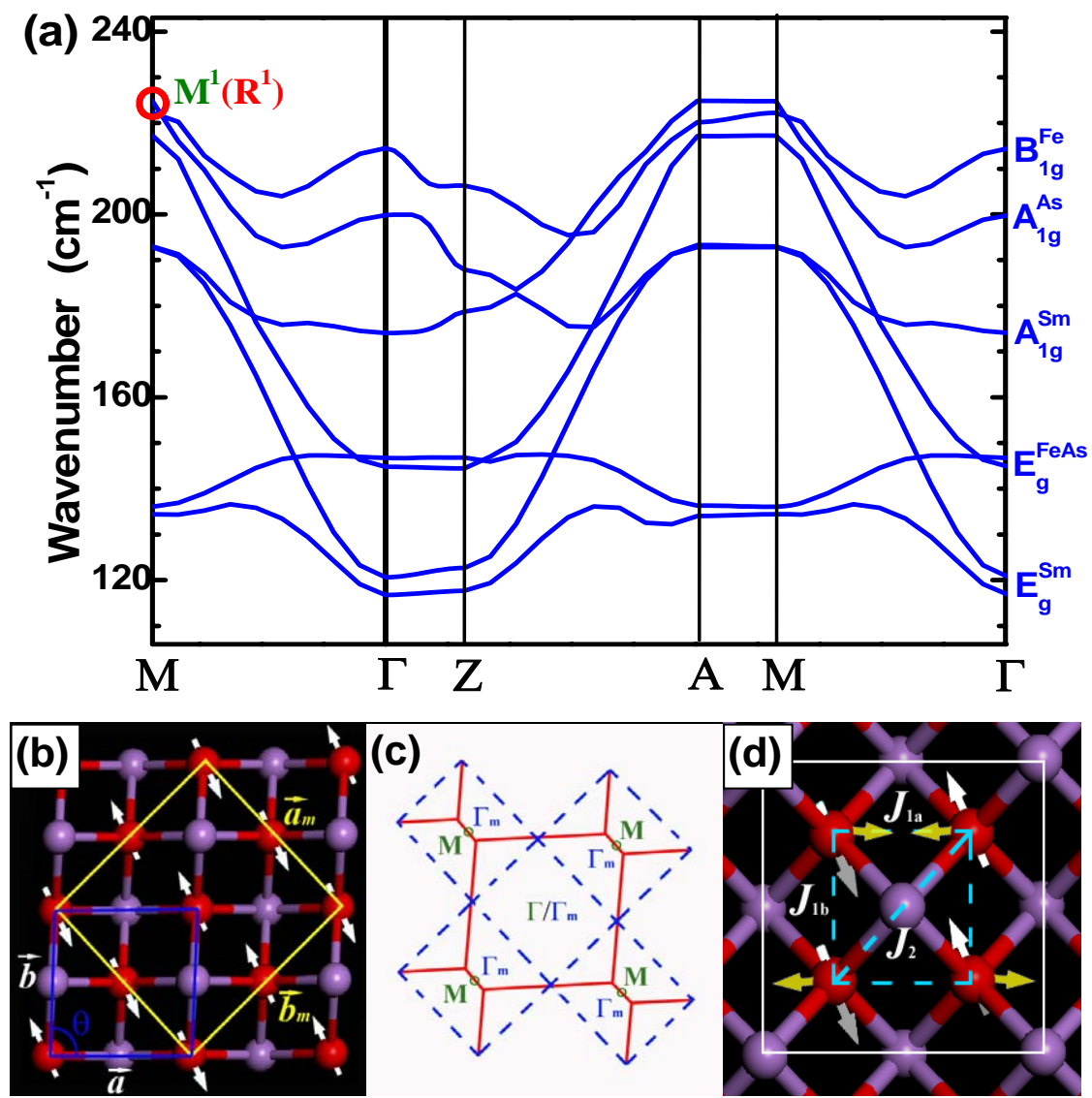

Table of contents

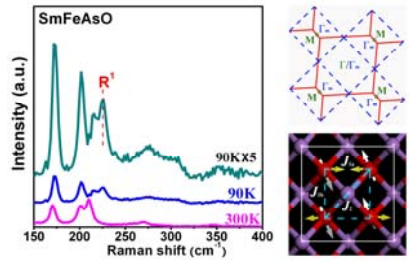